\documentclass[runningheads]{llncs}
\usepackage{graphicx}
\usepackage{amsmath}
\usepackage{amssymb}
\usepackage{enumerate}   
\usepackage{tikz}
\newtheorem{examp}{\sc Example}
\newtheorem{remk}{\sc Remark}
\newtheorem{corol}{\sc Corollary}
\newtheorem{defn}{\sc Definition}%
\newcommand{\bt}{\begin{theorem}}
\newcommand{\et}{\end{theorem}}
\newcommand{\bl}{\begin{lemma}}
\newcommand{\el}{\end{lemma}}
\newcommand{\bed}{\begin{defn}}
\newcommand{\eed}{\end{defn}}
\newcommand{\brem}{\begin{remk}}
\newcommand{\erem}{\end{remk}}
\newcommand{\bex}{\begin{examp}}
\newcommand{\eex}{\end{examp}}
\newcommand{\bcl}{\begin{corol}}
\newcommand{\ecl}{\end{corol}}

\numberwithin{equation}{section}
\numberwithin{theorem}{section}
\numberwithin{lemma}{section}
\begin{document}
\title{On Zero-Sum Two Person Perfect Information Semi-Markov Games\thanks{Supported by Department of Science and Technology, Govt. of India, INSPIRE Fellowship Scheme.}}
%
%
\author{Sagnik Sinha, Kushal Guha Bakshi \orcidID{0000-0002-5215-9828}}
%
%
\institute{Jadavpur University, Department Of Mathematics, 188, Raja S.C. Mallick Rd, Kolkata 700032\\
\email{sagnik62@gmail.com, kushalguhabakshi@gmail.com}\\
}
%
\maketitle
\begin{abstract}
 A zero-sum two-person  Perfect Information Semi-Markov game (PISMG) under limiting ratio 
average payoff has a value and both the maximiser and the minimiser have optimal pure semi-stationary strategies. We arrive at the result by first fixing an 
arbitrary initial state and forming the matrix of undiscounted payoffs corresponding to each pair of pure  stationary strategies of the two players and proving
that this matrix has a pure saddle point.

\end{abstract}
\section{Introduction}
A semi-Markov game (SMG) is a generalisation of a Stochastic (Markov) game (Shapley(1953) \cite{shapley1953stochastic}). Such games have already been studied in the literature 
(e.g. Lal- Sinha(1992) \cite{lal1992zero}, Luque-Vasquez(1999) \cite{luque1999semi}, Mondal(2015) \cite{mondal2015ordered}). Single player SMGs are called semi-Markov decision processes (SMDPs) which were introduced by Jewell(1963)  \cite{jewell1963markov} and Howard(1971) \cite{howard1971semi}. A perfect information semi-Markov game (PISMG) is a natural extension of perfect information stochastic games (PISGs) (Raghavan et al.(1997) \cite{thuijsman1997perfect}), where at each state all but one player is a dummy (i.e., he has only one available action in that state). Note that for such a game, perfect information is a state property. In this paper, we prove that such games (PISMGs) have a value and both players have pure semi-stationary optimal strategies under undiscounted (limiting ratio average) pay-offs. We prove this by showing the existence of a pure saddle point in the pay-off matrix of the game for each initial state. The paper is organized as follows. Section 2 contains definitions and properties of an undiscounted two person zero-sum semi-Markov game considered under limiting ratio average pay-off. Section 3 contains main result of this paper. In Section 4 we state the algorithm to compute a Cesaro limiting matrix of a transition matrix, proposed by Lazari et al., \cite{lazari2020new}. Section 5 contains a numerical example illustrating our theorem. Section 6 is reserved for the conclusion.

\section{Preliminaries}
\subsection{Finite zero-sum two-person  semi-Markov games}
A zero-sum two-person finite SMG is described by a collection of objects $\Gamma= <S, \lbrace A(s): s \in S \rbrace, \lbrace B(s) : s \in S \rbrace, q, P, r>$, where $S =\{1,2,\cdots,N\}$ is the finite non-empty state space and $A(s)=\lbrace 1,2,\cdots,m_s\rbrace, B(s)= \lbrace 1,2,\cdots,n_s\rbrace$ are respectively the non-empty sets of admissible actions of the players I and II respectively in the state $s$. Let us denote $K=\lbrace (s,i,j): s \in S, i \in A(s), j \in B(s) \rbrace$ to be the set of admissible triplets. For each $(s,i,j) \in K$, we denote $q(.\mid s,i,j)$ to be the transition law of the game. Given $(s,i,j) \in K$ and $s^{'} \in S$, let $\tau_{ij}^{ss^{'}}$ be the transition time random variable which denotes the time for a transition to a state $s^{'}$ from a state $s$ by a pair of actions $(i,j) \in A(s) \times B(s)$. Let $P_{ij}^{ss^{'}}= Prob (\tau_{ij}^{ss^{'}}\leq t)$ for each $(s,i,j)\in K, s^{'} \in S$ be a probability distribution function on [$0, \infty$) and it is called the conditional transition time distribution function. Finally $r$ is the real valued functions on $K$, which represents the immediate (expected) rewards for the player-I (and $-r$ is the immediate reward for player-II). Let us consider player I as the maximiser and player II as the minimiser in the zero-sum two person SMG.
The semi-Markov game over infinite time is played as follows. At the $1$st decision epoch, the game strats at $s_1 \in S$ and the players I and II simultaneously and independently choose actions $i_1 \in A(s_1)$ and $j_1 \in B(s_1)$ respectively. Consequently player I and II get immediate rewards $r(s_1,i_1,j_1)$ and $-r(s_1,i_1,j_1)$ respectively and the game moves to the state $s_2$ with probability $q(s_2\mid s_1,i_1,j_1)$. The sojourn time to move from state $s_1$ to the state $s_2$ is determined by the distribution function $P_{i_1j_1}^{s_1s_2}(.)$. After reaching the state $s_2$ on the next decision epoch, the game is repeated over infinite time with the state $s_1$ replaced by $s_2$.\\

By a strategy (behavioural) $\pi_1$ of the player I, we mean a sequence $\lbrace (\pi_1)_{n}(.\mid hist_{n}) \rbrace_{n=1}^{\infty}$, where $(\pi_1)_{n}$ specifies which action is to be chosen on the $n$-th decision epoch by associating with each history $hist_{n}$ of the system up to $n$th decision epoch (where $hist_{n}$=$(s_{1},a_{1},b_1,s_{2},a_{2},b_2 \cdots,s_{n-1}, a_{n-1},b_{n-1},\\s_{n})$ for $n \geq 2$, $hist_{1}=(s_{1})$ and $(s_{k}, a_{k}, j_k) \in K$  are respectively the state and actions of the players at the $k$-th decision epoch) a probability distribution $(\pi_1)_{n}(.\mid hist_{n})$ on $A(s_{n})$.  Behavioural strategy $\pi_2$ for player II can be defined analogously. Generally by any unspecified strategy, we mean behavioural strategy here. We denote $\Pi_1$ and $\Pi_2$ to be the sets of strategies (behavioural) of the players I and II respectively.\\
A strategy $f=\lbrace f_n\rbrace_{n=1}^{\infty}$ for the player I is called semi-Markov if for each $n$, $f_n$ depends on $s_1, s_n$ and the decision epoch number $n$. Similarly we can define a semi-Markov strategy $g=\{g_n\}_{n=1}^{\infty}$ for the player II.\\
A stationary strategy is a strategy that depends only on the current state. A stationary strategy for player I is defined as $N$ tuple $f=(f(1), f(2), \cdots,f(N))$, where each $f(s)$ is the probability distribution on $A(s)$ given by $f(s)=(f(s,1),f(s,2), \cdots, f(s,m_s))$. $f(s,i)$ denotes the probability of choosing action $i$ in the state $s$. By similar manner, one can define a stationary strategy $g$ for player II as $g=(g(1),g(2), \cdots, g(N))$ where each $g(s)$ is the probability distribution on $B(s)$. Let us denote $F_1$ and $F_2$ to be the set of stationary strategies for player I and II respectively.\\
A stationary strategy is called pure if any player selects a particular action with probability $1$ while visiting a state $s$. We denote $F_{1}^{s}$ and $F_{2}^{s}$ to be the set of pure stationary strategies of the players I and II respectively.\\
A semi-stationary strategy is a semi-Markov strategy which is independent of the decision epoch $n$, i.e., for a initial state $s_1$ and present state $s_2$, if a semi-Markov strategy $g(s_1,s_2,n)$ turns out to be independent of $n$, then we call it a semi-stationary strategy. Let $\xi_1$ and $\xi_2$ denote the set of semi-stationary strategies for the players I and II respectively and $\xi_{1}^{sp}$ and $\xi_{2}^{sp}$ denote the set of pure semi-stationary strategies for the players I and II respectively. \\
\textbf{Definition 1}
A zero-sum two person SMG $\Gamma= <S, \lbrace A(s): s \in S \rbrace, \lbrace B(s) : s \in S \rbrace, q, P, r>$ is called a perfect information semi-Markov game (PISMG) if the following properties hold\\
(i)$S=S_1 \cup S_2, S_1 \cap S_2 = \phi$.\\
(ii)$ \mid B(s)\mid =1$, for all $s \in S_1$, i.e., on $S_1$ player-II is a dummy.\\
(iii)$\mid A(s)\mid =1$, for all $s \in S_2$, i.e., on $S_2$ player-I is a dummy.\\
\subsection{ Zero-Sum Two-Person Semi-Markov Games under Limiting Ratio Average (Undiscounted) Payoff}
Let $(X_1,A_1,B_1,X_2,A_2,B_2\cdots)$ be a co-ordinate sequence in $S\times(A\times B \times S)^{\infty}$. Given behavioural strategy pair $(\pi_1,\pi_2) \in \Pi_1\times \Pi_2$, initial state $s \in S$, there exists a unique probability measure $P_{\pi_1 \pi_2}(. \mid X_1=s)$ (hence an expectation $E_{\pi_1 \pi_2}(. \mid X_1=s)$) on the product $\sigma$- field of  $S\times(A\times B \times S)^{\infty}$ by Kolmogorov's extension theorem. For a pair of strategies $(\pi_1, \pi_2) \in \Pi_1 \times \Pi_2$ for the players I and II respectively, the limiting ratio average (undiscounted) pay-off for player I, starting from a state $s \in S$ is defined by:
\begin{center}
	$\phi(s,\pi_1, \pi_2)$= $\liminf_{n \to \infty}\frac{E_{\pi_1 \pi_2} \sum_{m=1}^{n} [r(X_m,A_m,B_m) \mid X_1=s]} {E_{\pi_1 \pi_2} \sum_{m=1}^{n} [\bar{\tau}(X_m,A_m,B_m) \mid X_1=s]}$.	
\end{center}
Here $\bar{\tau}(s,i,j)= \sum_{s^{'} \in S} q(s^{'}\mid s,i,j)\int_{0}^{\infty}tdP_{ij}^{ss^{'}}(t)$ is the expected sojourn time in the state $s$ for a pair of actions $(i,j) \in A(s) \times B(s)$.\\
\textbf{Definition 2}
For each pair of stationary strategies $(f,g) \in F_1 \times F_2$ we define the transition probability matrix as $Q(f,g)= [q(s^{'} \mid s,f,g)]_{N \times N}$, where $q(s^{'}\mid s,f,g)=\sum_{i \in A(s)} \sum_{j \in B(s)} q(s^{'}\mid s,i,j) f(s,i)g(s,j)$ is the probability that starting from the state $s$, next state is $s^{'}$ when the players choose strategies $f$ and $g$ respectively (For a   stationary strategy $f$,  $f(s,i)$ denotes the probability of choosing action $i$ in the state $s$).

For any pair of stationary strategies $(f, g) \in F_1 \times F_2$ of player I and II, we write the undiscounted pay-off for player I as:\\
\begin{center}
	$\phi(s,f,g)= \liminf_{n \to \infty} \frac{\sum_{m=1}^n r^m(s,f,g)}
	{\sum_{m=1}^n\bar{\tau}^m(s,f,g)}$ for all $s \in S$.
\end{center}
Where $r^m(s,f,g)$ and $\bar{\tau}^m(s,f,g)$ are respectively the expected reward and expected sojourn time for player I at the $m$ th decision epoch, when player I chooses $f$ and player II chooses $g$ respectively and the initial state is $s$. We define $r(f,g)=[r(s,f,g)]_{N \times 1}$, $\bar{\tau}(f,g)= [\bar{\tau}(s,f,g)]_{N \times 1}$ and $\phi(f,g)=[\phi(s,f,g)]_{N \times 1}$ as expected reward, expected sojourn time and undiscounted pay-off vector for a pair of stationary strategy $(f,g) \in F_1\times F_2$. Now

$$\begin{array}{cc}
	r^{m}(s,f,g) ~~~~~= \sum_{s^{'} \in S} P_{fg}(X_m=s^{'} \mid X_1=s)r(s^{'},f,g)\\
	~~~~~~~~~~~~=\sum_{s^{'}\in S}r(s^{'},f,g)q^{m-1}(s^{'} \mid s,f,g)\\
	=[Q^{m-1}(f,g)r(f,g)](s)	
\end{array}$$
and
$$\begin{array}{cc}
	\bar{\tau}^{m}(s,f,g) ~~~~~= \sum_{s^{'} \in S} P_{fg}(X_m=s^{'} \mid X_1=s)\bar{\tau}(s^{'},f,g)\\
	~~~~~~~~~~~~=\sum_{s^{'}\in S}\bar{\tau}(s^{'},f,g)q^{m-1}(s^{'} \mid s,f,g)\\
	=[Q^{m-1}(f,g)\bar{\tau}(f,g)](s)	
\end{array}$$
Since $Q(f,g)$ is a Markov matrix, we have by Kemeny et al., \cite{kemeny1961finite}\\
\begin{center}
	$\lim_{n \to \infty} \frac{1}{n} \sum_{m=1}^{n}Q^m(f,g)$ exists and equals to $Q^{\ast}(f,g)$.
\end{center}
It is obvious that
\begin{center}
	$\lim_{n \to \infty}\frac{1}{n} \sum_{m=1}^{n}r^m(f,g)=[Q^{\ast}(f,g)r(f,g)](s)$
\end{center}
and
\begin{center}
	$\lim_{n \to \infty}\frac{1}{n} \sum_{m=1}^{n}\bar{\tau}^m(f,g)=[Q^{\ast}(f,g)\bar{\tau}(f,g)](s)$.
\end{center}
Thus we have for any pair of stationary strategies $(f_1,f_2) \in F_1 \times F_2$,
\begin{center}
	$\phi(s,f,g)=\frac{[Q^{*}(f,g)r(f,g)](s)}{[Q^*(f,g)\bar{\tau}(f,g)](s)}$ for all $s \in S$ 
\end{center}
where $Q^{*}(f,g)$ is the Cesaro limiting matrix of $Q(f,g)$.\\
\textbf{Definition 3}
A zero-sum two person undiscounted semi-Markov game is said to have a value vector $\phi=[\phi(s)]_{N \times 1}$ if $\sup_{\pi_1 \in \Pi_1}\inf_{\pi_2 \in \Pi_2} \phi(s,\pi_1,\pi_2)= \phi(s)= \inf_{\pi_2 \in \Pi_2} \sup_{\pi_1 \in \Pi_1} \phi(s,\pi_1,\pi_2)$ for all $s \in S$. A pair of strategies $(\pi_1^{\ast},\pi_2^{\ast}) \in \Pi_1,\times \Pi_2$ is said to be an optimal strategy pair for the players if $\phi(s,\pi_1^{\ast}, \pi_2) \geq \phi(s) \geq \phi(s, \pi_1,\pi_2^{\ast})$ for all $s \in S$ and all $(\pi_1,\pi_2) \in \Pi_1 \times \Pi_2$. Throughout this paper, we use the notion of undiscounted pay-off as limiting ratio average pay-off.
\section{Results}
\textbf{Theorem 1}\label{t1}
Any zero-sum two person undiscounted perfect information semi-Markov game has a solution in pure semi-stationary strategies under limiting ratio average pay-offs.
\begin{proof}
	Let $\Gamma=<S=S_{1} \cup S_{2}, A=\{A(s): s\in S_{1}\}, B=\{B(s): s \in S_{2}\}, q, P, r>$ be a zero-sum two person perfect information semi-Markov game under limiting ratio average pay-off, where $S=\{1,2,,\cdots,N\}$ is the finite state space. Let us fix an initial state $s \in S$. We assume that in $\mid S_{1}\mid$ number of states (i.e., states $\{1,2,\cdots,S_{1}\}$), player-II is a dummy and from states $\{ \mid S_{1} \mid +1, \cdots,\mid S_{1} \mid + \mid S_{2} \mid\}$ player-I is a dummy. We assume that in this perfect information game, player-I has $d_{1},d_{2},\cdots,d_{S_1}$ number of pure actions in the states where he is non-dummy and similarly player-II has $t_{S_{1}+1},t_{S_{1}+2},\cdots,t_{S_{1}+S_{2}}$ number of pure actions available in the states where he is non-dummy. Let $D_{1}= \Pi_{i=1}^{S_{1}}d_{i}$ and $D_{2}=\Pi_{i=S_{1}+1}^{S_{1}+S_{2}}t_{i}$. Let us the consider the pay-off matrix\\
	\[
	A_{D_{1} \times D_{2}} =
	\left[ {\begin{array}{cccc}
			\phi(s,f_1,g_1) & \phi(s,f_1,g_2) & \cdots & \phi(s,f_1,g_{D_{2}})\\
			\phi(s,f_2,g_1) & \phi(s,f_2,g_2) & \cdots & \phi(s,f_2,g_{D_{2}})\\
			\vdots & \vdots & \ddots & \vdots\\
			\phi(s,f_{D_{1}},g_1) & \phi(s,f_2,g_2) & \cdots & \phi(s,f_{D_{1}},g_{D_{2}})\\
	\end{array} } \right]
	\] Where $(f_1,f_2,\cdots,f_{D_{1}})$ and $(g_1,g_2,\cdots,g_{D_{2}})$ are the pure stationary strategies chosen by player-I and II repsectively. In order to prove the existence of a pure semi-stationary strategy, we have to prove that this matrix has a pure saddle point for each initial state $s \in S$. Now by theorem $2.1$ (``Some topics in two-person games'', in the Advances in Game Theory.(AM-52), Volume 52, 1964, page-$6$) proposed by Shapley \cite{dresher1964advances}, we know that, if A is the pay-off matrix of a two-person zero-sum game and if every $2\times 2$ submatrix of $A$ has a saddle point, then A has a saddle point. So, we concentrate only on a $2\times2$ matrix and observe if it has a saddle point or not. We consider the $2\times2$ submatrix:\\
	\begin{center}
		$\left[ {\begin{array}{cccc}
				\phi(s,f_i,g_j) & \phi(s,f_i,g_{j^{'}}) \\
				\phi(s,f_{i^{'}},g_j) & \phi(s,f_i^{'},g_{j^{'}}) \\
		\end{array} } \right]$
		
	\end{center}
	Where $i^{'}, i \in \{d_{1}, d_{2}\cdots, d_{S_1}\}, (i\neq i^{'})$ and $j,j^{'} \in \{t_{S_{1}+1},t_{S_{1}+2},\cdots,t_{S_{1}+S_{2}}\}, (j\neq j^{'})$. Now, by suitably renumbering the strategies, we can write the above sub-matrix as:\\
	\begin{center}
		\[
		A^{'}_{2 \times 2} =
		\left[ {\begin{array}{cccc}
				\phi(s,f_1,g_1) & \phi(s,f_1,g_2) \\
				\phi(s,f_2,g_1) & \phi(s,f_2,g_2) \\
		\end{array} } \right]
		\]
	\end{center} Now we know\\
	\begin{center}
		$\phi(s,f_{i.},g_{.j})=\frac{\sum_{t=1}^{S_{1}}[q^{\ast}(t \mid s, f_{i.}) r(t, f_{i.})] + \sum_{v=S_{1}+1}^{S_{1}+S_{2}}[q^{\ast}(v \mid s, g_{.j}) r(v, g_{.j})]} {\sum_{t=1}^{S_{1}}[q^{\ast}(t \mid s, f_{i.}) \tau(t, f_{i.})] + \sum_{v=S_{1}+1}^{S_{1}+S_{2}}[q^{\ast}(v \mid s, g_{.j}) \tau(v, g_{.j})]}$
	\end{center}
	We replace $\phi(s,f_{i.},g_{.j})$ by the expression above in the matrix $A$. Let us rename the elements of the $2\times2$ sub-matrix as we consider the following two cases when $A$ can not have a pure saddle point.\\
	Case-1: $\phi(s,f_1,g_1)$ is row-minimum and column-minimum, $\phi(s,f_1,g_2)$ is row-maximum and column-maximum, $\phi(s,f_2,g_1)$ is row-maximum and column-maximum and $\phi(s,f_2,g_2)$ is row-minimum and column-minimum. These four conditions can be written as:\\ $\phi(s,f_1,g_1)<\phi(s,f_1,g_2)$, $\phi(s,f_1,g_1)<\phi(s,f_2,g_1)$ $\phi(s,f_2,g_2)<\phi(s,f_2,g_1)$ and $\phi(s,f_2,g_2)<\phi(s,f_1,g_2)$.
	So, the above four inequalities can be written elaborately as:\\
	\begin{eqnarray}
		\frac{\sum_{t=1}^{S_{1}}[q^{\ast}(t \mid s, f_{1.}) r(t, f_{1.})] + \sum_{v=S_{1}+1}^{S_{1}+S_{2}}[q^{\ast}(v \mid s, g_{.1}) r(v, g_{.1})]} {\sum_{t=1}^{S_{1}}[q^{\ast}(t \mid s, f_{1.}) \tau(t, f_{1.})] + \sum_{v=S_{1}+1}^{S_{1}+S_{2}}[q^{\ast}(v \mid s, g_{.1}) \tau(v, g_{.1})]}\\\nonumber< \frac{\sum_{t=1}^{S_{1}}[q^{\ast}(t \mid s, f_{1.}) r(t, f_{1.})] + \sum_{v=S_{1}+1}^{S_{1}+S_{2}}[q^{\ast}(v \mid s, g_{.2}) r(v, g_{.2})]} {\sum_{t=1}^{S_{1}}[q^{\ast}(t \mid s, f_{1.}) \tau(t, f_{1.})] + \sum_{v=S_{1}+1}^{S_{1}+S_{2}}[q^{\ast}(v \mid s, g_{.2}) \tau(v, g_{.2})]}\\
		\noindent
		\frac{\sum_{t=1}^{S_{1}}[q^{\ast}(t \mid s, f_{1.}) r(t, f_{1.})] + \sum_{v=S_{1}+1}^{S_{1}+S_{2}}[q^{\ast}(v \mid s, g_{.1}) r(v, g_{.1})]} {\sum_{t=1}^{S_{1}}[q^{\ast}(t \mid s, f_{1.}) \tau(t, f_{1.})] + \sum_{v=S_{1}+1}^{S_{1}+S_{2}}[q^{\ast}(v \mid s, g_{.1}) \tau(v, g_{.1})]}\\\nonumber<
		\frac{\sum_{t=1}^{S_{1}}[q^{\ast}(t \mid s, f_{2.}) r(t, f_{2.})] + \sum_{v=S_{1}+1}^{S_{1}+S_{2}}[q^{\ast}(v \mid s, g_{.1}) r(v, g_{.1})]} {\sum_{t=1}^{S_{1}}[q^{\ast}(t \mid s, f_{2.}) \tau(t, f_{2.})] + \sum_{v=S_{1}+1}^{S_{1}+S_{2}}[q^{\ast}(v \mid s, g_{.1}) \tau(v, g_{.1})]}\\
		\noindent
		\frac{\sum_{t=1}^{S_{1}}[q^{\ast}(t \mid s, f_{2.}) r(t, f_{2.})] + \sum_{v=S_{1}+1}^{S_{1}+S_{2}}[q^{\ast}(v \mid s, g_{.2}) r(v, g_{.2})]} {\sum_{t=1}^{S_{1}}[q^{\ast}(t \mid s, f_{2.}) \tau(t, f_{2.})] + \sum_{v=S_{1}+1}^{S_{1}+S_{2}}[q^{\ast}(v \mid s, g_{.2}) \tau(v, g_{.2})]}\\\nonumber<
		\frac{\sum_{t=1}^{S_{1}}[q^{\ast}(t \mid s, f_{2.}) r(t, f_{2.})] + \sum_{v=S_{1}+1}^{S_{1}+S_{2}}[q^{\ast}(v \mid s, g_{.1}) r(v, g_{.1})]} {\sum_{t=1}^{S_{1}}[q^{\ast}(t \mid s, f_{2.}) \tau(t, f_{2.})] + \sum_{v=S_{1}+1}^{S_{1}+S_{2}}[q^{\ast}(v \mid s, g_{.1}) \tau(v, g_{.1})]}\\
		\noindent
		\frac{\sum_{t=1}^{S_{1}}[q^{\ast}(t \mid s, f_{2.}) r(t, f_{2.})] + \sum_{v=S_{1}+1}^{S_{1}+S_{2}}[q^{\ast}(v \mid s, g_{.2}) r(v, g_{.2})]} {\sum_{t=1}^{S_{1}}[q^{\ast}(t \mid s, f_{2.}) \tau(t, f_{2.})] + \sum_{v=S_{1}+1}^{S_{1}+S_{2}}[q^{\ast}(v \mid s, g_{.2}) \tau(v, g_{.2})]}\\ \nonumber< 
		\frac{\sum_{t=1}^{S_{1}}[q^{\ast}(t \mid s, f_{1.}) r(t, f_{1.})] + \sum_{v=S_{1}+1}^{S_{1}+S_{2}}[q^{\ast}(v \mid s, g_{.2}) r(v, g_{.2})]} {\sum_{t=1}^{S_{1}}[q^{\ast}(t \mid s, f_{1.}) \tau(t, f_{1.})] + \sum_{v=S_{1}+1}^{S_{1}+S_{2}}[q^{\ast}(v \mid s, g_{.2}) \tau(v, g_{.2})]} 
	\end{eqnarray}
	We rename the strategies $f_{1.}, f_{2.},g_{.1}$ and $g_{.2}$ as $1.$, $2.$, $.1$ and $.2$ respectively to avoid notational complexity. Hence, $(3.1)$ yields\\
	
	\begin{equation}
\begin{split}
\sum_{t=1}^{S_1}\sum_{v=S_{1}+1}^{S_1 + S_2}q^{\ast}(t\mid s,1.)q^{\ast}(v\mid s,.2)[\tau(t,1.)r(v,.2)-r(t,1.)\tau(v,.2)]\\+ \sum_{t=1}^{S_{1}}\sum_{v=S_{1}+1}^{S_1+S_2}q^{\ast}(t \mid s, 1.)q^{\ast}(v \mid s, .1)[r(t,1.)\tau(v,.1)\\-r(v,.1)\tau(t,1.)]
\sum_{v=S_{1}+1}^{S_1+S_2}\sum_{v=S_{1}+1}^{S_1+S_2}q^{\ast}(v \mid s,.1)[\tau(v,.1)r(v,.2)-r(v,.1)\tau(v,.2)]\textgreater0
\end{split}
	\end{equation}

$(3.3)$ yields\\
\begin{equation}
\begin{split}
\sum_{t=1}^{S_1}\sum_{v=S_{1}+1}^{S_1 + S_2}q^{\ast}(t\mid s,2.)q^{\ast}(v\mid s,.1)[\tau(t,2.)r(v,.1)-r(t,2.)\tau(v,.1)]\\+ \sum_{v=S_{1}+1}^{S_1+S_2}\sum_{v=S_{1}+1}^{S_1+S_2}q^{\ast}(v \mid s,.2)q^{\ast}(v \mid s,.1)[\tau(v,.2)r(v,.1)\\-r(v,.2)\tau(v,.1)]+ \sum_{t=1}^{S_{1}}\sum_{v=S_{1}+1}^{S_1+S_2}q^{\ast}(t\mid s, 2.)q^{\ast}(v\mid,s .2)[r(t,2.)\tau(v,.2)-\tau(t,.2)r(v,.2)]>0
\end{split}
\end{equation}
	
	$(3.2)$ yields\\
	\begin{equation}
		\begin{split}
			\sum_{t=1}^{S_1}\sum_{t=1}^{S_1}q^{\ast}(t\mid s,1.)q^{\ast}(t\mid s,.2)[\tau(t,1.)r(t,2.)-r(t,1.)\tau(t,2.)]\\+
			\sum_{t=1}^{S_{1}}\sum_{v=S_{1}+1}^{S_1+S_2}q^{\ast}(t \mid s, 1.)q^{\ast}(v \mid s, .1)[r(v,1.)\tau(t,.1)\\-r(t,1.)\tau(v,.1)]+
			\sum_{t=1}^{S_1}\sum_{v=S_{1}+1}^{S_1+S_2}q^{\ast}(t\mid s,2.)q^{\ast}(v \mid s,.1)[\tau(v,.1)r(t,2.)-r(v,.1)\tau(t,2.)]>0	
		\end{split}
	\end{equation}
	
	$(3.4)$ yields\\
	\begin{equation}
		\begin{split}
			\sum_{t=1}^{S_1}\sum_{t=1}^{S_1}q^{\ast}(t\mid s,1.)q^{\ast}(t\mid s,.2)[r(t,1.)\tau(t,2.)-r(t,2.)\tau(t,1.)]\\+
			\sum_{t=1}^{S_{1}}\sum_{v=S_{1}+1}^{S_1+S_2}q^{\ast}(t \mid s, 2.)q^{\ast}(v \mid s, .2)[r(v,.2)\tau(t,2.)\\-r(t,2.)\tau(v,.2)]+ \sum_{t=1}^{S_1}\sum_{v=S_{1}+1}^{S_1+S_2}q^{\ast}(t \mid s,1.)q^{\ast}(v \mid s,.2)[r(t,1.)\tau(v,.2)-r(v,.2)\tau(t,1.)]>0		
		\end{split}
	\end{equation}

	Using the fact that, $0<q^{\ast}(s^{'}\mid s, a)<1$, (where $s,s^{'} \in \{ 1,2,\cdots,N\}$, $a$ is the action chosen by either player-I or II) and adding $(3.5)$ and $(3.6)$, we get\\
	\begin{equation}
		\begin{split}
			\sum_{t=1}^{S_1}\sum_{v=S_{1}+1}^{S_1+S_2} (\tau(t,1.)r(v,.2)-r(t,1.)\tau(v,.2))+
			\sum_{t=1}^{S_1}\sum_{v=S_{1}+1}^{S_1+S_2} (r(t,1.) \tau(v,.1)-\tau(t,1.)r(v,.1))+\\
			\sum_{t=1}^{S_1}\sum_{v=S_{1}+1}^{S_1+S_2} (r(v,.1)\tau(t,2.)-r(t,2.)\tau(v,.1))+
			\sum_{t=1}^{S_1}\sum_{v=S_{1}+1}^{S_1+S_2}\\ (r(t,.2)\tau(v,2.)-\tau(t,2.)r(v,.2))>0
		\end{split}
	\end{equation}
	
	Similarly adding $(3.7)$ and $(3.8)$, we get\\
	\begin{equation}
		\begin{split}
			\sum_{t=1}^{S_1}\sum_{v=S_{1}+1}^{S_1+S_2} (\tau(v,.2)r(t,1.)-r(v,.2)\tau(t,1.))+
			\sum_{t=1}^{S_1}\sum_{v=S_{1}+1}^{S_1+S_2} (r(v,.1) \tau(t,1.)-\tau(v,.1)r(t,1.))+\\
			\sum_{t=1}^{S_1}\sum_{v=S_{1}+1}^{S_1+S_2} (r(t,2.)\tau(v,.1)-r(v,.1)\tau(t,2.))+
			\sum_{t=1}^{S_1}\sum_{v=S_{1}+1}^{S_1+S_2}\\ (r(v,.2)\tau(t,2.)-\tau(t,.2)r(v,2.))>0
		\end{split}
	\end{equation}
	From $(3.9)$ and $(3.10)$ we clearly get a contradiction. Now we consider the next case:
	
	Case-2: $\phi(s,f_1,g_1)$ is row maximum and column maximum, $\phi(s,f_1,g_2)$ is row minimum and column minimum, $\phi(s,f_2,g_1)$ is row-minimum and column minimum and $\phi(s,f_2,g_2)$ is row-maximum and column-maximum. These four conditions can be written as: $\phi(s,f_1,g_1)>\phi(s,f_1,g_2)$, $\phi(s,f_1,g_1)>\phi(s,f_2,g_1)$, $\phi(s,f_2,g_2)>\phi(s,f_2,g_1)$ and $\phi(s,f_2,g_2)>\phi(s,f_1,g_2)$. We can re-write them as follows:
	\begin{eqnarray}
		\frac{\sum_{t=1}^{S_{1}}[q^{\ast}(t \mid s, f_{1.}) r(t, f_{1.})] + \sum_{v=S_{1}+1}^{S_{1}+S_{2}}[q^{\ast}(v \mid s, g_{.1}) r(v, g_{.1})]} {\sum_{t=1}^{S_{1}}[q^{\ast}(t \mid s, f_{1.}) \tau(t, f_{1.})] + \sum_{v=S_{1}+1}^{S_{1}+S_{2}}[q^{\ast}(v \mid s, g_{.1}) \tau(v, g_{.1})]}\\\nonumber> \frac{\sum_{t=1}^{S_{1}}[q^{\ast}(t \mid s, f_{1.}) r(t, f_{1.})] + \sum_{v=S_{1}+1}^{S_{1}+S_{2}}[q^{\ast}(v \mid s, g_{.2}) r(v, g_{.2})]} {\sum_{t=1}^{S_{1}}[q^{\ast}(t \mid s, f_{1.}) \tau(t, f_{1.})] + \sum_{v=S_{1}+1}^{S_{1}+S_{2}}[q^{\ast}(v \mid s, g_{.2}) \tau(v, g_{.2})]}.\\
		\noindent
		\frac{\sum_{t=1}^{S_{1}}[q^{\ast}(t \mid s, f_{1.}) r(t, f_{1.})] + \sum_{v=S_{1}+1}^{S_{1}+S_{2}}[q^{\ast}(v \mid s, g_{.1}) r(v, g_{.1})]} {\sum_{t=1}^{S_{1}}[q^{\ast}(t \mid s, f_{1.}) \tau(t, f_{1.})] + \sum_{v=S_{1}+1}^{S_{1}+S_{2}}[q^{\ast}(v \mid s, g_{.1}) \tau(v, g_{.1})]}\\\nonumber>
		\frac{\sum_{t=1}^{S_{1}}[q^{\ast}(t \mid s, f_{2.}) r(t, f_{2.})] + \sum_{v=S_{1}+1}^{S_{1}+S_{2}}[q^{\ast}(v \mid s, g_{.1}) r(v, g_{.1})]} {\sum_{t=1}^{S_{1}}[q^{\ast}(t \mid s, f_{2.}) \tau(t, f_{2.})] + \sum_{v=S_{1}+1}^{S_{1}+S_{2}}[q^{\ast}(v \mid s, g_{.1}) \tau(v, g_{.1})]}.\\
		\noindent
		\frac{\sum_{t=1}^{S_{1}}[q^{\ast}(t \mid s, f_{2.}) r(t, f_{2.})] + \sum_{v=S_{1}+1}^{S_{1}+S_{2}}[q^{\ast}(v \mid s, g_{.2}) r(v, g_{.2})]} {\sum_{t=1}^{S_{1}}[q^{\ast}(t \mid s, f_{2.}) \tau(t, f_{2.})] + \sum_{v=S_{1}+1}^{S_{1}+S_{2}}[q^{\ast}(v \mid s, g_{.2}) \tau(v, g_{.2})]}\\\nonumber>
		\frac{\sum_{t=1}^{S_{1}}[q^{\ast}(t \mid s, f_{2.}) r(t, f_{2.})] + \sum_{v=S_{1}+1}^{S_{1}+S_{2}}[q^{\ast}(v \mid s, g_{.1}) r(v, g_{.1})]} {\sum_{t=1}^{S_{1}}[q^{\ast}(t \mid s, f_{2.}) \tau(t, f_{2.})] + \sum_{v=S_{1}+1}^{S_{1}+S_{2}}[q^{\ast}(v \mid s, g_{.1}) \tau(v, g_{.1})]}.\\
		\noindent
		\frac{\sum_{t=1}^{S_{1}}[q^{\ast}(t \mid s, f_{2.}) r(t, f_{2.})] + \sum_{v=S_{1}+1}^{S_{1}+S_{2}}[q^{\ast}(v \mid s, g_{.2}) r(v, g_{.2})]} {\sum_{t=1}^{S_{1}}[q^{\ast}(t \mid s, f_{2.}) \tau(t, f_{2.})] + \sum_{v=S_{1}+1}^{S_{1}+S_{2}}[q^{\ast}(v \mid s, g_{.2}) \tau(v, g_{.2})]}\\ \nonumber> 
		\frac{\sum_{t=1}^{S_{1}}[q^{\ast}(t \mid s, f_{1.}) r(t, f_{1.})] + \sum_{v=S_{1}+1}^{S_{1}+S_{2}}[q^{\ast}(v \mid s, g_{.2}) r(v, g_{.2})]} {\sum_{t=1}^{S_{1}}[q^{\ast}(t \mid s, f_{1.}) \tau(t, f_{1.})] + \sum_{v=S_{1}+1}^{S_{1}+S_{2}}[q^{\ast}(v \mid s, g_{.2}) \tau(v, g_{.2})]}. 
	\end{eqnarray}
	Like the previous case we also rename the strategies $f_{1.}, f_{2.},g_{1.}$ and $g_{2.}$ as $1.$, $2.$, $.1$ and $.2$ respectively to avoid notational complexity. Hence, $(3.11)$ yields\\
	\begin{equation}
		\begin{split}
			\sum_{t=1}^{S_{1}}q^{\ast}(t\mid s, 1.)\tau(t,1.)\sum_{v=S_{1}+1}^{S_1+S_2}q^{\ast}(v\mid,s .1)r(v,.1)+ \sum_{t=1}^{S_{1}}q^{\ast}(t\mid s, 1.)r(t,1.)\sum_{v=S_{1}+1}^{S_1+S_2}q^{\ast}(v\mid,s .2)\tau(v,.2)\\+ \sum_{v=S_{1}+1}^{S_1+S_2}q^{\ast}(v \mid s,.2)r(v,.1)\sum_{v=S_{1}+1}^{S_1+S_2}q^{\ast}(v \mid s,.2)\tau(v,.2)-\sum_{t=1}^{S_1}\sum_{v=S_{1}+1}^{S_1 + S_2}q^{\ast}(t\mid s,1.)q^{\ast}(v\mid s,.2)\tau(t,1.)r(v,.2)\\-\sum_{t=1}^{S_{1}}\sum_{v=S_{1}+1}^{S_1+S_2}q^{\ast}(t \mid s, 1.)q^{\ast}(v \mid s, .1)r(t,1.)\tau(v,.1)- \sum_{v=S_{1}+1}^{S_1+S_2}q^{\ast}(v \mid s,.2)r(v,.2)\sum_{v=S_{1}+1}^{S_1+S_2}q^{\ast}(v \mid s,.1)\tau(v,.1)>0
		\end{split}
	\end{equation}
	$(3.13)$ yields
	
	\begin{equation}
		\begin{split}
			\sum_{t=1}^{S_{1}}\sum_{v=S_{1}+1}^{S_1+S_2}q^{\ast}(t\mid s, 2.)\tau(t,2.)q^{\ast}(v\mid,s .2)r(v,.2)+\sum_{t=1}^{S_{1}}\sum_{v=S_{1}+1}^{S_1+S_2}q^{\ast}(t\mid s, 2.)r(t,2.)q^{\ast}(v\mid,s .1)\tau(v,.1)\\+\sum_{v=S_{1}+1}^{S_1+S_2}\sum_{v=S_{1}+1}^{S_1+S_2}q^{\ast}(v \mid s,.1)\tau(v,.1)q^{\ast}(v \mid s,.2)r(v,.2)-\sum_{t=1}^{S_1}\sum_{v=S_{1}+1}^{S_1 + S_2}q^{\ast}(t\mid s,2.)q^{\ast}(v\mid s,.2)\tau(t,2.)r(v,.1)\\-
			\sum_{t=1}^{S_{1}}\sum_{v=S_{1}+1}^{S_1+S_2}q^{\ast}(t \mid s, 1.)q^{\ast}(v \mid s, .1)\tau(t,.2)r(v,2.)- \sum_{v=S_{1}+1}^{S_1+S_2}\sum_{v=S_{1}+1}^{S_1+S_2}q^{\ast}(v \mid s,.2)\tau(v,.2)q^{\ast}(v \mid s,.1)r(v,.1)>0		
		\end{split}
	\end{equation}
	$(3.12)$ yields\\
	\begin{equation}
		\begin{split}	
			\sum_{t=1}^{S_{1}}q^{\ast}(t\mid s, 2.)\tau(t,2.)\sum_{v=S_{1}+1}^{S_1+S_2}q^{\ast}(v\mid,s .1)r(v,.1)+ \sum_{t=1}^{S_{1}}q^{\ast}(t\mid s, 1.)r(t,1.)\sum_{v=S_{1}+1}^{S_1+S_2}q^{\ast}(v\mid,s .1)\tau(v,.1)\\ +\sum_{t=1}^{S_1}q^{\ast}(t \mid s,1.)r(t,1.)\sum_{t=1}^{S_1}q^{\ast}(t \mid s,2.)\tau(t,2.)- \sum_{t=1}^{S_1}\sum_{t=1}^{S_1}q^{\ast}(t\mid s,1.)q^{\ast}(t\mid s,.2)\tau(t,1.)r(t,2.)- \\ \sum_{t=1}^{S_{1}}\sum_{v=S_{1}+1}^{S_1+S_2}q^{\ast}(t \mid s, 1.)q^{\ast}(v \mid s,.1)r(v,.1)\tau(t,1.)- \sum_{t=1}^{S_1}q^{\ast}(t \mid s,.2)r(t,.2)\sum_{v=S_{1}+1}^{S_1+S_2}q^{\ast}(v \mid s,.1)\tau(v,.1)>0
		\end{split}
	\end{equation}
	$(3.14)$ yields\\
	\begin{equation}
		\begin{split}
			\sum_{t=1}^{S_{1}}q^{\ast}(t\mid s, 1.)\tau(t,1.)\sum_{v=S_{1}+1}^{S_1+S_2}q^{\ast}(v\mid,s .2)r(v,.2)+ \sum_{t=1}^{S_{1}}q^{\ast}(t\mid s, 2.)r(t,2.)\sum_{v=S_{1}+1}^{S_1+S_2}q^{\ast}(v\mid,s .2)\tau(v,.2)\\ +\sum_{t=1}^{S_1}q^{\ast}(t \mid s,2.)r(t,2.)\sum_{t=1}^{S_1}q^{\ast}(t \mid s,1.)\tau(t,1.)-\sum_{t=1}^{S_1}\sum_{t=1}^{S_1}q^{\ast}(t\mid s,1.)q^{\ast}(t\mid s,.2)r(t,1.)\tau(t,2.)\\ -\sum_{t=1}^{S_{1}}\sum_{v=S_{1}+1}^{S_1+S_2}q^{\ast}(t \mid s, 2.)q^{\ast}(v \mid s, .2)r(v,.2)\tau(t,2.)- \sum_{t=1}^{S_1}q^{\ast}(t \mid s,1.)r(t,1.)\sum_{v=S_{1}+1}^{S_1+S_2}q^{\ast}(v \mid s,.2)\tau(v,.2)>0 		
		\end{split}
	\end{equation}
	Similarly using the fact that $0<q^{\ast}(s^{'}\mid s, a)<1$, (where $s,s^{'} \in \{ 1,2,\cdots,N\}$, $a$ is the action chosen by either player-I or II) and adding $(3.15)$ and $(3.16)$, we get\\
	\begin{equation}
		\begin{split}
			\sum_{t=1}^{S_1}\sum_{v=S_{1}+1}^{S_1+S_2} (\tau(v,.2)r(t,1.)-r(v,.2)\tau(t,1.))+
			\sum_{t=1}^{S_1}\sum_{v=S_{1}+1}^{S_1+S_2} (r(v,.1) \tau(t,1.)-\tau(v,.1)r(t,1.))+\\
			\sum_{t=1}^{S_1}\sum_{v=S_{1}+1}^{S_1+S_2} (r(t,2.)\tau(v,.1)-r(v,.1)\tau(t,2.))+
			\sum_{t=1}^{S_1}\sum_{v=S_{1}+1}^{S_1+S_2}\\ (r(v,.2)\tau(t,2.)-r(t,.2)\tau(v,2.))>0
		\end{split}
	\end{equation}
	Now adding $(3.17)$ and $(3.18)$ we get\\
	\begin{equation}
		\begin{split}
			\sum_{t=1}^{S_1}\sum_{v=S_{1}+1}^{S_1+S_2} (\tau(t,1.)r(v,.2)-r(t,1.)\tau(v,.2))+
			\sum_{t=1}^{S_1}\sum_{v=S_{1}+1}^{S_1+S_2} (r(t,1.) \tau(v,.1)-\tau(t,1.)r(v,.1))+\\
			\sum_{t=1}^{S_1}\sum_{v=S_{1}+1}^{S_1+S_2} (r(v,.1)\tau(t,2.)-r(t,2.)\tau(v,.1))+
			\sum_{t=1}^{S_1}\sum_{v=S_{1}+1}^{S_1+S_2}\\ (r(t,.2)\tau(v,2.)-r(v,.2)\tau(t,2.))>0
		\end{split}
	\end{equation}
	From $(3.19)$ and $(3.20)$ we get a contradiction. Thus, every $2\times2$ submatrix has a pure saddle point and by theorem $2.1$ proposed by Shapley (\cite{dresher1964advances}, (page-$6$)), the matrix $A$ has a pure saddle point and the game $\Gamma$ has a pure 
	stationary optimal strategy pair for each initial state. Suppose $(f_{1},f_{2},\cdots,f_{N})$ be optimal pure stationary strategies for player-I when the initial states are $1,2,\cdots,N$ respectively and $(g_{1},g_{2},\cdots,g_{N})$ be optimal pure stationary strategies for player-II when the initial states are $1,2,\cdots,N$ respectively. The $f^{\ast}=(f_{1},f_{2},\cdots,f_{N})$ and $g^{\ast}=(g_{1},g_{2},\cdots,g_{N})$ are the optimal pure semi-stationary strategies for player-I and II respectively in the perfect information semi-Markov game $\Gamma$.
\end{proof}

\section{Calculating the Cesaro Limiting Matrix of A Transition Matrix}

Lazari et al., \cite{lazari2020new} proposed an algorithm to compute the Cesaro limiting matrix of any Transition (Stochastic) matrix $Q$ with $n$ states. The algorithm runs as follows:
\begin{flushleft}
	Input: Let  the transition matrix $Q \in M_n(\mathbb{R})$ (where $M_n(\mathbb{R})$ is the set of $n \times n$ matrices over the field of real numbers).
	
	Output: The Cesaro limiting matrix $Q^{\ast} \in M_n(\mathbb{R})$.\\
	Step $1$: Determine the characteristic polynomial $C_{Q}(z)= \mid Q-zI_{n}\mid$.\\
	Step $2$: Divide the polynomial $C_{Q}(z)$ by $(z-1)^{m(1)}$ (where $m(1)$ is the algebraic multiplicity of the eigenvalue $z_{0}=1$) and call it quotient $T(z)$.\\
	Step $3$: Compute the quotient matrix $W=T(Q)$.\\
	Step $4$: Determine the limiting matrix $Q^{\ast}$ by dividing the matrix $W$ by the sum of its elements of any arbitrary row.
\end{flushleft}

\section{An Example}
\textbf{Example:} Consider a PISMG $\Gamma$ with four states $S=\{1,2,3,4\}$, $A(1)=\{1,2\}=A(2)$, $B(1)=B(2)=\{ 1\}$, $B(3)=B(4)=\{1,2\}$, $A(3)=A(4)=\{1\}.$ Player II is the dummy player in the state $1$ and $2$ and player I is the dummy player for the states $3$ and $4$. Rewards, transition probabilities and expected sojourn times for the players are given below
\vskip 1 cm
State-1: \begin{tabular}{|r|}
	\hline
	1.1\\ ($\frac{1}{2}$,$\frac{1}{2}$,0,0)\\  1\\
	\hline
	1\\ ($\frac{1}{3}$,$\frac{2}{3}$,0,0)\\ 0.9\\
	\hline
\end{tabular}
State-2: \begin{tabular}{|r|}
	\hline
	3.1\\ ($\frac{1}{2}$,$\frac{1}{2}$,0,0)\\ 1\\
	\hline
	3\\ ($\frac{2}{3}$,$\frac{1}{3}$,0,0)\\ 1.1\\
	\hline
\end{tabular}
State-3: \begin{tabular}{|c|c|}
	\hline
	$3$ & $5.8$\\
	$(0,0,1,0)$ & ($0,0,1,0$)\\
	1 &  2\\
	\hline
\end{tabular}\\
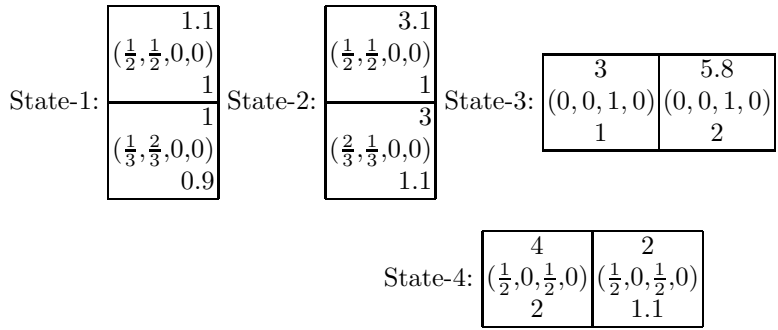
\begin{center}
	State-4: \begin{tabular}{|c|c|}
		\hline
		$4$ & $2$ \\
		($\frac{1}{2}$,0,$\frac{1}{2}$,0) & ($\frac{1}{2}$,0,$\frac{1}{2}$,0)\\
		$2$ & $1.1$\\
		\hline
	\end{tabular}
\end{center}
Where a cell \begin{tabular}{|r|}
	\hline
	$(r)$\\$(q_1,q_2,q_3,q_4)$\\$\bar{\tau}$\\
	\hline
\end{tabular} represents that $r$ is the immediate rewards of the playes, $q_1$, $q_2$, $q_3$, $q_4$ represents that the next states are $1$, $2$, $3$ and $4$ respectively and $\bar{\tau}$ is the expected sojourn time if this cell is chosen at present. Here player I is the row player and player II is the column player. Player-I has the pure statioanry strategies $f_{1}=\{(1,0),(1,0),1,1\}$, $f_{2}=\{(1,0),(0,1),1,1\}$, $f_{3}=\{(0,1),(1,0),1,1\}$ and $f_{4}=\{(0,1),(0,1),1,1\}$. Similarly the pure stationary strategies for player-II are $g_{1}=\{1,1,(1,0),(1,0)\}$, $g_{2}=\{1,1,(1,0),(0,1)\}$, $g_{3}=\{1,1,(0,1),(1,0)\}$ and $g_{4}=\{1,1,(0,1),(0,1)\}$.  Now, we calculate the undiscounted value of the PISMG for each initial state by complete enumeration method. Using the alogrithm described in section $4$, we calculate the Cesaro limiting matrices as follows:\\
$Q^{\ast}(f_1,g_{1})=Q^{\ast}(f_2,g_{1})=Q^{\ast}(f_3,g_{1})=Q^{\ast}(f_4,g_{1})= 
\left[\begin{array}{rrrrr} 
	\frac{1}{2} & \frac{1}{2} & 0 & 0\\
	\frac{1}{2} & \frac{1}{2} & 0 & 0\\
	0 & 0 & 1 & 0\\
	\frac{1}{2} & \frac{1}{2} & 0 & 0\\
\end{array}\right]$,
$Q^{\ast}(f_2,g_{1})=Q^{\ast}(f_{2},g_{2})=Q^{\ast}(f_{2},g_{3}) =Q^{\ast}(f_2,g_{4})=
\left[\begin{array}{rrrrr} 
	\frac{1}{2} & \frac{1}{2} & 0 & 0\\
	\frac{2}{3} & \frac{1}{3} & 0 & 0\\
	0 & 0 & 1 & 0\\
	\frac{1}{2} & 0 & \frac{1}{2} & 0\\
\end{array}\right]$,
$Q^{\ast}(f_{3},g_{1})=Q^{\ast}(f_{3},g_{2})=Q^{\ast}(f_{3},g_{3})=Q^{\ast}(f_{3},g_{4})=
\left[\begin{array}{rrrrr} 
	\frac{1}{3} & \frac{2}{3} & 0 & 0\\
	\frac{1}{2} & \frac{1}{2} & 0 & 0\\
	0 & 0 & 1 & 0\\
	\frac{1}{2} & 0 & \frac{1}{2} & 0\\
\end{array}\right]$, $Q^{\ast}(f_{4},g_{1})=Q^{\ast}(f_{4},g_{2})=Q^{\ast}(f_{4},g_{3})=Q^{\ast}(f_{4},g_{4})= \left[\begin{array}{rrrrr} 
	\frac{1}{3} & \frac{2}{3} & 0 & 0\\
	\frac{2}{3} & \frac{1}{3} & 0 & 0\\
	0 & 0 & 1 & 0\\
	\frac{1}{2} & 0 & \frac{1}{2} & 0\\
\end{array}\right]$. Now the reward vector $\hat{r}(f_1,g_1)=(1.1,3.1,3,4)$ and expected sojourn time vector $\bar{\tau}(f_1)=(1,1,1,2)$. Thus by using the definition of $\hat{\phi}$, we get $\hat{\phi}(f_1,g_1)=(2.1, 2.1, 3, 0.9)$. Similarly we calculate the undiscounted pay-offs for other pairs of pure stationary strategies as: $\hat{\phi}(f_1,g_2)=(2.1, 2.1, 3, 0.9)$, 
$\hat{\phi}(f_1,g_3)=(2.1, 2, 2.9, 0.9)$, 
$\hat{\phi}(f_1,g_4)=(2.1, 2.1, 2.9, 0.9)$,
$\hat{\phi}(f_2,g_1)=(1.8353, 1.8362, 3, 0.53)$, $\hat{\phi}(f_2,g_3)=(1.8353 ,1.8362, 2.9, 1.2776)$, $\hat{\phi}(f_2,g_2)\\=(1.8353, 1.8362, 3, 0.58)$, $\hat{\phi}(f_2,g_4)=(1.8353,1.8362, 2.9, 1.2773)$, $\hat{\phi}(f_3,g_1)=(2.2985, 2.2985, 3, 0.4088)$, $\hat{\phi}(f_{3},g_2)=(2.2985, 2.2985, 3, 0.4088)$, $\hat{\phi}(f_3,g_3)=(2.2985, 2.2985, 2.9, 0.4088)$,
$\hat{\phi}(f_3,g_4)=(2.2985, 2.2988,\\ 2.9, 1.2182)$,
$\hat{\phi}(f_4,g_1)=(2.2979, 2.2985, 3, 0.4277)$,
$\hat{\phi}(f_4,g_3)=(2.112, 2.1129, 2.9, 1.2141)$,
$\hat{\phi}(f_4,g_2)=(2.112, 2.1129, 3, 0.4267)$,
$\hat{\phi}(f_4,g_4)=(2.112, 2.1129, 2.9, 1.2137)$.
For initial state $1$, we get the pay-off matrix $A$ as described in section $3$, as:\\ 
\[
A^{1}_{4 \times 4 } =
\left[ {\begin{array}{cccc}
		2.1 & 2.1 & 2.1 & 2.1\\
		1.8353 & 1.8353 & 1.8353 & 1.8353\\
		2.2985 & 2.2985 & 2.2985 & 2.2985\\
		2.2979 & 2.112 & 2.112 & 2.112\\
\end{array} } \right].
\]
So, this matrix has a pure saddle point at the $3$rd  row, $3$rd column position and we conclude $(f_3,g_3)$ is the optimal pure stationary strategy pair for the players for initial state $1$. Similarly for initial state $2$, $3$ and $4$ we get pay-off matrices as:\\
$A^{2}_{4 \times 4 } =
\left[ {\begin{array}{cccc}
		2.1 & 2.1 & 2 & 2.1\\
		1.8362 & 1.8362 & 1.8362 & 1.8362\\
		2.2985 & 2.2985 & 2.2985 & 2.2988\\
		2.2985 & 2.1129 & 2.1129 & 2.1129\\
\end{array} } \right]$,
$A^{3}_{4 \times 4 } =
\left[ {\begin{array}{cccc}
		3 & 3 & 2.9 & 2.9\\
		3 & 3 & 2.9 & 2.9\\
		3 & 3 & 2.9 & 2.9\\
		3 & 3 & 2.9 & 2.9\\
\end{array} } \right]$ and
$A^{4}_{4 \times 4 } =
\left[ {\begin{array}{cccc}
		0.9 & 0.9 & 0.9 & 0.9\\
		0.53 & 0.58 & 1.2776 & 1.2773\\
		0.4088 & 0.4088 & 0.4088 & 1.2182\\
		0.4277 & 0.4267 & 1.2141 & 1.2137\\
\end{array} } \right]$. The optimal pure statioanry strategy pairs of the player-I and player-II for the initial states $2$, $3$ and $4$ are $(f_{3},g_{3})$, $(f_{1},g_3)$ and $(f_1,g_2)$ repectively. Thus the optimal pure semi-stationary strategy is $f^{\ast}=(f_3,f_3,f_1,f_1)$ and $g^{\ast}=(g_3,g_3,g_3,g_2)$ for the players I and II respectively and the game has a value $(2.2985,2.2985,2.9,0.9)$. 
\section{Conclusion}
The purpose of this paper is to show that there exists an optimal pure semi-stationary strategy pair by just looking at the pay-off matrix in any Perfect Information 
semi-Markov game. Thus, the existence of the value and a pair of pure semi-stationary optimals for the players in a zero-sum two person Perfect Information undiscounted 
semi-Markov game can be obtained as a corollary of Shapley's paper (1964) (\cite{dresher1964advances}) directly without going through the discounted version. Furthermore, 
the existence of a pure optimal strategy (not necessarily stationary/ semi-statioanry) for an $N$ person Perfect Information non-cooperative semi-Markov game under 
any standard (discounted/ undiscounted) pay-off criteria can be shown. We shall elaborate on this in a forthcoming paper.

\end{document}